\documentclass[twocolumn,secnumarabic,amssymb, nobibnotes, aps, prd]{revtex4-2}

\usepackage{hyperref}
\usepackage{mathtools,amsthm}

\usepackage[dvipsnames]{xcolor}

\usepackage{graphicx}
\graphicspath{ {./Figures/} }
\usepackage{float}

\newtheorem{theorem}{Theorem}[section]

\begin{document}


\title{Half-life Measurements of Highly Charged Radioisotopes by Nuclear Recoil in a Penning Trap}


\author{Scott Moroch}
\email[]{smoroch@mit.edu}
\affiliation{University of Maryland, College Park, Department of Materials Science and Engineering}
\author{Carolyn Chun}
\affiliation{University of Maryland, College Park, Department of Materials Science and Engineering}
\author{Doug VanDerwerken}
\affiliation{United States Naval Academy}
\author{Ariana Shearin}
\affiliation{University of Maryland, College Park, Department of Materials Science and Engineering}
\author{Brian L. Beaudoin}
\affiliation{University of Maryland, College Park, Department of Materials Science and Engineering}
\author{Klaus Blaum}
\affiliation{Max-Planck-Institut f\"ur Kernphysik, Saupfercheckweg 1, 69117 Heidelberg, Germany}
\author{Timothy W. Koeth}
\affiliation{University of Maryland, College Park, Department of Materials Science and Engineering}


\date{\today}

\begin{abstract}
We present a novel method for measuring the half-life of highly charged radioisotopes by non-destructive nuclear recoil detection in a Penning ion trap. A specific emphasis is placed on $\rm ^7Be^{3+}$,  which plays a crucial role in stellar evolution and the production of solar neutrinos. 
The determination of the half-life is necessary to constrain the free electron capture rate in the solar environment, but is difficult to measure by existing techniques.
Simulations of the sympathetic cooling of the recoiled daughter nuclei ($\rm ^7Li^{3+}$) with the trapped cloud of $\rm ^7Be^{3+}$ demonstrate a decay detection efficiency of $99.5\%$. A statistical analysis of half-life measurements on ensembles containing hundreds of ions shows that a final statistical uncertainty of {less than} $5\%$ is achieved with only 500 measured decays. 
By coherent control of hyperfine populations in trapped ions, the fidelity of the technique we describe enables the direct measurement and manipulation of state-dependent decay branching ratios for the first time.

\end{abstract}


\maketitle

\section{Introduction}

\indent In radioactive decay by electron capture (EC) a proton-rich nucleus captures a bound-state electron $p+e^{-} \rightarrow n + \nu_e$,  when beta decay is energetically forbidden ($ Q<2m_e$). 
This decay mechanism is a result of the overlap of the electron wavefunction with that of the nucleus. {Segre proposed} in 1947 that manipulating the electron wavefunction in these isotopes could result in a measurable variation to the half-life  \cite{SegreMinutes}.  \

\indent $\rm ^{7}Be$ is the lowest-Z isotope to decay exclusively by electron capture with a neutral half-life of $53$ days. Experiments over the last several decades have  successfully measured 
half-life changes in $\rm ^{7}Be$ by implanting the isotope in chemical structures that provide variations in the electron density at the nucleus resulting in 
variations in the half-life of a few percent \cite{Segre_7BeVariation,7Be_hostMaterials,7Be_Metals, 7Be_Fullerenes, C60_cages}. A more direct method of varying the electron wavefunction is to generate highly charged ions resulting in half-life variation orders of magnitude larger than previously measured \cite{Patyk_paper,Seigen_Paper}.

\indent The Schottky Mass Spectrometry (SMS) technique \cite{Schottky} has been used to perform 
half-life measurements of highly charged ions for many electron-capture and beta-decay radioisotopes in storage rings \cite{Litvinov2023}. These experiments have probed the effect of atomic structure on decay rates; in some cases enabling the observation of rare decay modes \cite{BoundStateBeta, Nature_205Tl, First_Bound_State,PhysRevLett.133.022502}. The technique relies on measuring the small difference in cyclotron frequencies of mother and daughter ions in a storage ring. At a radioactive ion beam facility large ion ensembles ($\sim 10^6$ ions) can be used to achieve sufficient decay statistics, despite limitations in the ion storage time ($\sim$ 5-30 mins), due to ion losses through collisions with background molecules. 

We present a novel application of cryogenic Penning ion traps to measure decay rate perturbations in highly charged radioisotopes, with $\rm ^{7}Be^{3+}$ as an example. For this case, electron recombination rates and a half-life much longer than storage times make the measurement difficult to perform \cite{ISOLDE_ring, Litvinov2023}. This particular ion of this isotope is of interest due to its critical role in the proton-proton chain in stellar evolution \cite{Bahcall_Solar_Models,Bahcall_7Be}. We show that uncertainties comparable to storage ring techniques are achievable in a table-top apparatus by measuring decays in a small ion ensemble ($<10^3$ ions) on an event-by-event basis by detecting nuclear recoil.  

The two primary contributions of this paper are as follows. First, we provide the simulations of a nuclear recoil-based method for measuring decay rate changes in a Penning ion trap with near $100\%$ detection fidelity.  Second, we give the statistical analysis of half-life measurements for small ensembles ($<10^{3}$) of ions. The motivation for studying $\rm ^7Be$ and its general decay scheme are presented in Section \ref{Motivation}. In Section \ref{Half-Life Measurements} we describe the Penning ion trap and the use of non-destructive image-current detection for measuring nuclear decay on an event-by-event basis. In Section \ref{Statistical Analysis}, we present the statistical analysis of the half-life measurements in small ensembles, including a discussion of systematics. We demonstrate that a statistical uncertainty of less than 5\% can be achieved with 500 measured decays. 
In Section \ref{Experimental Realization}, we present the technical details of a practical experiment. {In Section \ref{Hyperfine section}, we demonstrate that the technique enables the direct measurement of decay branching ratios from different hyperfine states through kinematic reconstruction of the daughter ion.} Finally, in Section \ref{Discussion}, we discuss other interesting applications and extensions of the technique. 

\section{Motivation and $\rm ^7Be$ Decay\label{Motivation}}

\begin{figure}[h]
\centering
\includegraphics[width=0.38\textwidth]{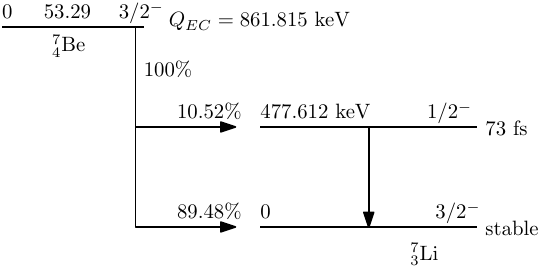}
\caption{In the $\rm ^7Be$ electron capture decay scheme, approximately 90\% of decays occur directly to the ground state of $\rm^7Li$. The remaining 10\% decay first to an excited state of $\rm^7Li$, which emits a 477 keV $\gamma$-ray in $73$ fs.}
\label{fig:decay scheme}
\end{figure}

\indent $\rm^7Be$ is the lowest-Z isotope to decay exclusively by electron capture ($Q=862$ keV), with a neutral half-life of 53 days and a stable $\rm^7Li$ daughter isotope \cite{7Be_Decay_info}. The general decay scheme is presented in Figure \ref{fig:decay scheme}. Approximately 90\% of decays occur via a Fermi {or Gamow-Teller} transition to the ground state of $\rm^7Li$. The remaining 10\% proceed via a Gamow-Teller transition to an excited state of $\rm^7Li$ that subsequently decays in $73$ fs to the ground state by emission of a 477 keV $\gamma$-ray. The long half-life and large decay Q-value make $\rm ^7Be$ the suitable candidate for a demonstration of the Penning trap technique. 

These measurements are motivated in part by a long-standing interest to improve our understanding of the influence of the electronic structure on the nuclear half-life in electron-capture radioisotopes. A list of experiments measuring the $\rm^7Be$ half-life under different environmental conditions and material structures can be found in \cite{7Be_Science,NORMAN200115,Ray_Science,PhysRevC.73.034323}.
Additionally, as theoretically described in \cite{Folan} and discussed in Section \ref{sec: atomic structure}, in hydrogen-like systems the hyperfine splitting in the ground state gives rise to states of different total angular momentum. 
Conservation of angular momentum constrains the decay branching ratios (in this case to either $\rm^7Li^*$ or $\rm ^7Li$) from hyperfine populations. Studies of this has been conducted in heavy nuclei where the ions are all populated in the ground hyperfine state (assuming no re-population in the ring) \cite{140Pr, 142Pm, 122I}. The simple structure of $\rm ^7Be$ and the ability to produce highly charged ions makes it a very attractive system for these studies as well. As we discuss in Section \ref{Hyperfine section}, the use of well-developed techniques to coherently control hyperfine states in a Penning ion trap, coupled with the detection scheme presented in this paper, provides a new avenue to explore these effects.

The specific interest in $\rm ^7Be$ also arises from its critical role in the proton-proton chain in stellar evolution and the production of solar neutrinos \cite{Bahcall_Solar_Models}. The abundance of $\rm ^7Be$ in the sun is determined by its destruction through

electron and proton capture, $\rm  ^7Be(e^-,\nu)^7Li$ and $\rm ^7Be(p,\gamma)^8B$, and its production in $\rm ^3He(\alpha,\gamma)^7Be$ \cite{Bahcall_7Be}. In the hot solar environment, all ionization states of $\rm ^7Be$ are present \cite{osti_4567269, Gruzinov_1997}. The destruction rates of $\rm ^7Be$ determine the branching ratios of the later proton-proton chains, and subsequently determine the $\rm ^7Be$ and $\rm ^8B$ solar neutrino fluxes \cite{osti_2429322}. 
Precision measurements of the solar neutrino flux (such as those in \cite{Borexino}) coupled with $\rm ^7Be$ decay rates from bound electrons would provide valuable insight to the free electron capture decay rate in a stellar environment, which presently cannot be measured in the lab. 


\subsection{Decay Recoil Energy \label{Decay Recoil Energy Sec}}

\begin{figure}[h]
\centering
\includegraphics[width=0.42\textwidth]{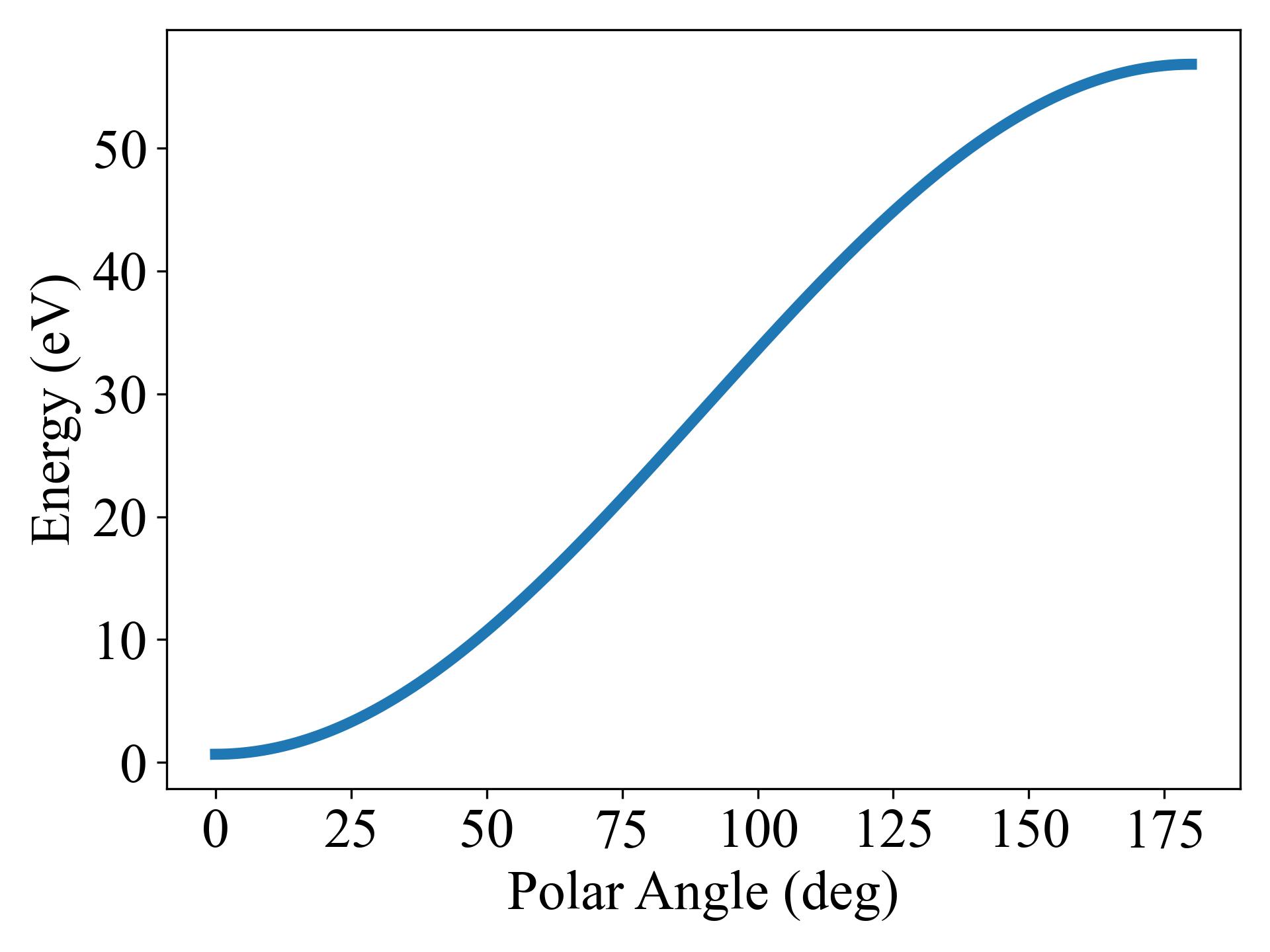}
\caption{Energy distribution of the final ground-state $\rm ^7Li$ as a function angle of gamma emission angle.}
\label{fig:excited state recoil}
\end{figure}

$\rm ^7Be$ electron capture decay is a two-body decay, resulting in the daughter nucleus and a neutrino. Neglecting the neutrino mass, the recoil kinetic energy of the daughter nucleus is given as follows,

\begin{equation}
    T_d = \frac{Q^2}{2 \left (Q+m_dc^2 \right )}.
\end{equation}
Here, $Q$ is the energy of the decay (862 keV in the case of $\rm ^7Be$) and $ m_d$ is the mass of the daughter nucleus. Nearly 90\% of $\rm ^7Be$ decays occur to the ground state of $\rm ^7Li$. In this case the $\rm ^7Li$ has a mono-energetic recoil energy of 56.83 eV. The remaining 10\% of decays are to $\rm ^7Li^{*}$, where a 477 keV $\gamma$-ray is then isotropically ejected. This results in a range of recoil energies from 0.66 to 56.83 eV, dependent on the direction of emission, as shown in Figure \ref{fig:excited state recoil}. Because the gamma ray is emitted isotropically from the isomer, the probability density of emitting into a particular polar angle ($\theta$) is:

\begin{equation}
    \rm P_{\gamma}(\theta) = \frac{1}{2} \sin(\theta).
    \label{eq: polar angle PDF}
\end{equation}

Taking into account Equation \ref{eq: polar angle PDF} and Figure \ref{fig:excited state recoil}, gives a
probability density function (PDF) for the energy that is
approximately uniform (see Appendix \ref{appB} for details).

\subsection{$\rm ^{7}Be^{3+}$ Atomic Structure} \label{sec: atomic structure}

In hydrogen-like $\rm ^{7}Be^{3+}$, the ground state electronic configuration is $\rm ^1S_{1/2}$. $\rm ^{7}Be$ has a nuclear spin of $I=\frac{3}{2}$, resulting in two hyperfine states with total angular momentum $F=1$ and $F=2$, and energy splitting of $\Delta_{HF} = 2 \pi* 30.4 $ GHz. Because the magnetic moment of $\rm ^{7}Be$ is negative, $\mu =-1.398(15)\mu_{N}$, the $F=2$ sub-level is the ground state \cite{7Be_magnetic_moment}. The ions in this experiment are produced by electron impact ionization (discussed in Section \ref{Experimental Realization}), at a temperature much greater than the hyperfine splitting ($k_BT>>\Delta_{HF}$). Therefore, $\rm ^{7}Be^{3+}$ ions are produced with the {hyperfine} states populated according to their statistical weights, which simulates the
population in a stellar environment. Additionally the lifetime of a hyperfine state scales as $Z^{-9}$, where $Z$ is the atomic number \cite{Sobelman, Patyk_paper}. Thus for $\rm ^{7}Be^{3+}$ the lifetime of the upper $F=1$ state is $\mathcal{O}({10^2})$ years and it can, therefore, be assumed that the hyperfine populations of the ensemble are fixed after production. 

\begin{figure}[H]
\centering
\includegraphics[width=0.45\textwidth]{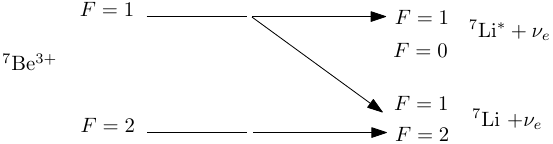}
\caption{$\rm ^7Be^{3+}$ decay from hyperfine states}
\label{fig:hyperfine decay scheme}
\end{figure}

The possible decay modes from each hyperfine state must conserve total angular momentum which restricts the possible decay modes. The ground state of $\rm ^7Li$ has nuclear spin $I=\frac{3}{2}$ and the electron neutrino has spin $s = \frac{1}{2}$, thus the possible angular momenta are $F=2$ and $F=1$. However, the isomeric state ($\rm ^7Li^*$) has nuclear spin $I = \frac{1}{2}$, giving rise to angular momenta $F=1$ and $F=0$. Consequently, the lower hyperfine state in $\rm ^7Be^{3+}$ can decay only to the ground state of $\rm ^7Li$. The allowed decay modes are shown in Figure \ref{fig:hyperfine decay scheme}. These effects were first predicted in \cite{Folan}, and to date have not been measured in $\rm ^7Be$.  

\section{Half-life Measurements by Nuclear Recoil in a Penning Trap\label{Half-Life Measurements}} 

\subsection{Penning Ion Trap} \label{Penning trap details} \indent A Penning trap uses a strong, axial uniform magnetic field ($\vec{B} = B \hat{z}$) and a weak, hyperbolic electrostatic potential, to confine ions.  The electrostatic potential, $U(r, z)$ is given by

\begin{equation} \label{potential}
    {U(r,z) = \frac{C_2U_0}{2d^2} (2z^2-r^2)}
\end{equation}
where $ C_2$ is an expansion coefficient, $ U_0$ is the voltage applied to the trap, and $d$ is the characteristic trap size~\cite{Geonium}. This potential is formed by a set of hyperbolic electrodes or with a set of five or seven cylindrical electrodes {(for example, see \cite{GABRIELSE1989319})}. The ion's motion can be decomposed into three simple harmonic eigenmotions: one axially ($\hat{z}$) and two in the radial plane: a fast oscillation about the magnetic field lines (modified cyclotron) and a slow {$\vec{E} \times \vec{B}$} drift about the center of the trap (magnetron). These three motions have corresponding frequencies $\omega_z$, $\omega_+$ and $\omega_-$, respectively. Typical values for $\rm ^7Be^{3+}$ with $B=3$ T, $d = 7$ mm and $U_0=25 $ V are: $\omega_z = 2\pi \times 537 $ kHz, $\omega_+ = 2\pi \times 19.6$ MHz and $\omega_- = 2\pi \times 7.3$ kHz. 

\subsection{Ion Cooling and Detection}

An ion {of charge $q$ and mass $m$} oscillating in the trap will induce a small image current ({$\sim 10$ fA for an ion at 4 K in the Measurement Trap described below}) in the electrodes used to form the potential in Equation~[\ref{potential}] \cite{Shockley}. The trap electrodes can be connected to a high-Q tank circuit (resonator) that, when tuned to the resonant frequency of the ion in the trap, presents a large effective parallel resistance $R_p$ and converts the current to a measurable voltage \cite{ULMER2013}. The image current is dissipated through this resistance as heat, that reduces the energy of the ion with a time constant \cite{Wayne_M_Itano_1995, Bolometric},

\begin{equation} \label{cooling rate}
    {\tau_{res} = \frac{mD_{eff}^2}{R_pq^2}}.
\end{equation}
{where $D_{eff}$ defines an effective distance between the ion and the electrode \cite{Geonium}}. An ion in equilibrium with the resonator ($ T_{ion} = T_{res} = 4$ K) shorts the noise spectrum and produces a well-defined dip~\cite{Wineland}. Critically, for a small number of ions (10s), this dip width scales linearly with the number of stored ions. This linearity can be exploited to accurately count the initial number of trapped ions~\cite{Antiproton_Lifetime}, and will be discussed in Section \ref{Ensemble Preparation}. For an ion with energy much greater than the resonator ($T_{ion} >> T_{res}$) the signal appears as a peak in the noise spectrum~\cite{ULMER2013}. The technique of peak detection enables the detection of a hot recoil ion, which in the case of $\rm ^7Be$ is at minimum 0.66 eV. The recoil ion is therefore a factor of at least 2000 higher in temperature than that of the detection circuit. 

\begin{figure*}[t]
\centering
\includegraphics[width=0.7\textwidth]{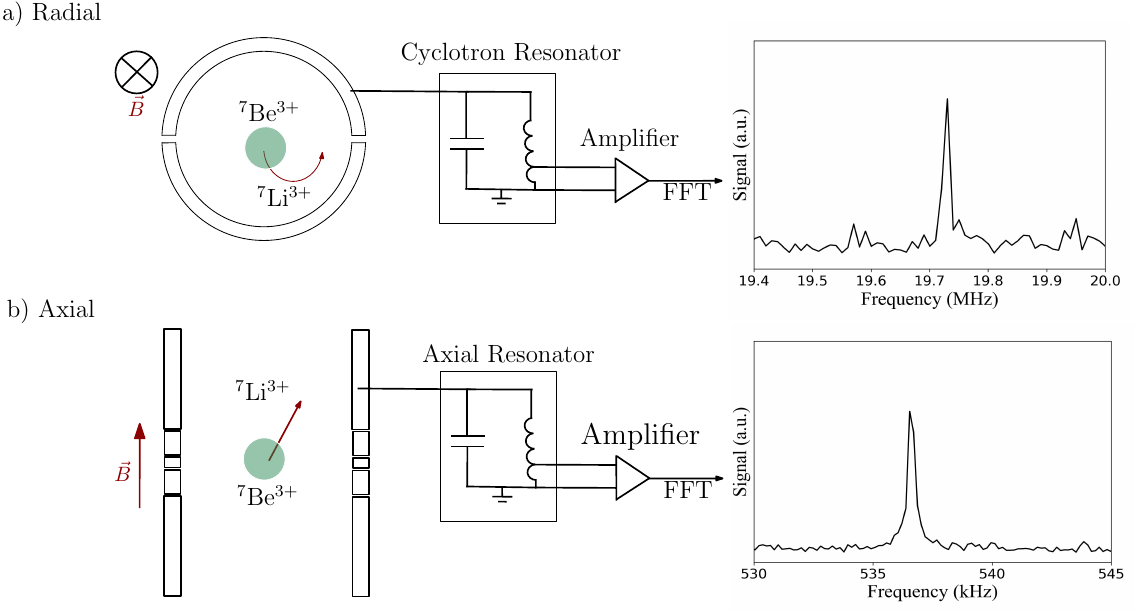}
\caption{Non-destructive detection scheme and SIMION simulation of the induced image current signal for the radial (a) and axial (b) modes of a single non-interacting ion.}
\label{fig:detection scheme}
\end{figure*}

\subsection{Recoil in Penning Trap: Single Particle Picture}

We propose trapping a known number of ions ($ N_0$). When an ion decays it recoils with an energy between 0.66 and 56.83 eV (see Section \ref{Decay Recoil Energy Sec}), depending on the decay {path}. As the ion oscillates in the trap, this energy will then be transferred to two cryogenic resonant circuits, coupled to the axial and {modified} cyclotron modes, respectively. With an energy above the noise floor of the detector, the ion signal can be resolved on an event-by-event basis. Since the measurement relies on the signature of a nuclear recoil rather than the relative abundance of different mass states ($\rm ^7Li$ vs. $\rm^7Be$), it is effectively background-free. Measurements of the energy in the axial and radial modes in the Penning trap enables a kinematic reconstruction of the decay that can provide direct information about the decay branching ratios.  

In the Penning trap the instantaneous kinetic and potential energies can be decomposed according to the three modes: axial, modified cyclotron, and magnetron. For an ion decay at the center of the trap, the average energy in the axial and radial modes is 

\begin{equation}
    E_z = \frac{q a_z^2 C_2 U_0}{d^2}
\end{equation}

\begin{equation}
    {E_\pm=\frac{\rho_\pm^2 m \left( \omega_\pm^2 -\omega_z^2/2 \right)}{2 d^2}}
\label{eq: radial energies}
\end{equation}
where $a_z$ and $\rho_\pm$ are the axial and radial amplitudes respectively. For the conditions of the initial ensemble in this experiment, $\rho_+ \omega_+/\rho_- \omega_-$ is much greater than unity, and thus, by Equation~[\ref{eq: radial energies}], $|E_+|>>|E_-|$. Therefore, in the analysis we assume all of the energy in the radial plane couples to the {modified} cyclotron mode. A recoil ion with total energy $E$ will have an energy coupled into axial and radial modes $ E_z=E \cos^2(\theta)$ and $ E_+=E \sin^2(\theta)$, where $\theta$ is measured with respect to the magnetic axis of the trap.

In a single particle picture (one that ignores the ion cloud) the ability to resolve a decay depends exclusively on the energy of the particle with respect to the noise floor of the detector. As shown in Section \ref{Decay Recoil Energy Sec}, the minimum recoil energy (0.66 eV) arises when $\rm ^7Be$ decays to $\rm ^7Li^*$ and the subsequent $\gamma$ is emitted in the direction of the momentum of the $\rm ^7Li^*$. As this energy is still far above the noise floor of the detector, it is expected that the detection of the decays through the induced image current will be sensitive to $100 \%$ of decays. A simulation of the induced image current signal from a single recoiled ion in the Penning trap is shown in Figure \ref{fig:detection scheme}, along with a pictorial representation of the detection scheme. The ion trajectory and image current is simulated in SIMION, an ion optics program \cite{SIMION}.

\subsection{Ion Recoil: Space Charge Effects} \label{Space Charge Section}

The motion of the recoiled ion will be damped by sympathetic cooling with the cold $\rm ^7Be^{3+}$ ions. To detect a single decay it is critical that this sympathetic cooling time is longer than the time needed to resolve the decayed ion on the resonator. These interactions will set a lower bound on the detectable recoil energy and an upper bound on the number of ions stored during a given measurement. 

\subsubsection{Sympathetic Cooling Time}

Consider a cloud of 500 ions in thermal equilibrium with a resonator at 4 K in the axial and {modified} cyclotron modes {(henceforth referred to as simply the axial and cyclotron detectors)}. When a single ion decays the daughter nucleus begins oscillating with a large amplitude in both the axial and radial modes and interacts with the cold cloud of ions through Coulomb collisions. The hot ion will ultimately cool to the equilibrium temperature. 

The cold cloud of ions may be treated as a sphere of radius $R$ with $N_0$ ions of charge $q$. The equilibrium radius of the cloud depends on the trapping parameters and we estimate it to be $R \approx 200 $ $ \space \mu m$. The number density of ions is estimated to be $n=3 \times 10^{13} $ $\rm  m^{-3}$. Inside of this cloud the hot ion engages in Coulomb collisions with a cooling (Spitzer) time constant, $ \tau_{Coul}$, given by

\begin{equation} \label{Coulomb}
    \tau_{Coul.} = \frac{4 \sqrt{2} \pi \epsilon_0^2 E^{3/2}\sqrt{m}}{nq^4 ln \Lambda} 
\end{equation}
where $\epsilon_0$ is the permittivity of free space, $E$ is the energy of the recoil ion, $m$ is the mass of the recoil ion and $ ln(\Lambda)$ is known as the Coulomb Logarithm ($\approx 6$ for the parameters given here)~\cite{Spitzer,Freidberg}. However, because of the ion's large oscillation amplitude, it spends only a fraction of its axial ($F_z$) and radial ($F_+$) oscillations interacting inside of the cold cloud. The total sympathetic cooling time for the hot ion is, consequently, increased to

\begin{equation} \label{time_Symp}
    \tau_{Symp} = \frac{1}{F_z F_+} \tau_{Coul.}
\end{equation}.

\begin{figure}[H]
\centering
\includegraphics[width=0.45\textwidth]{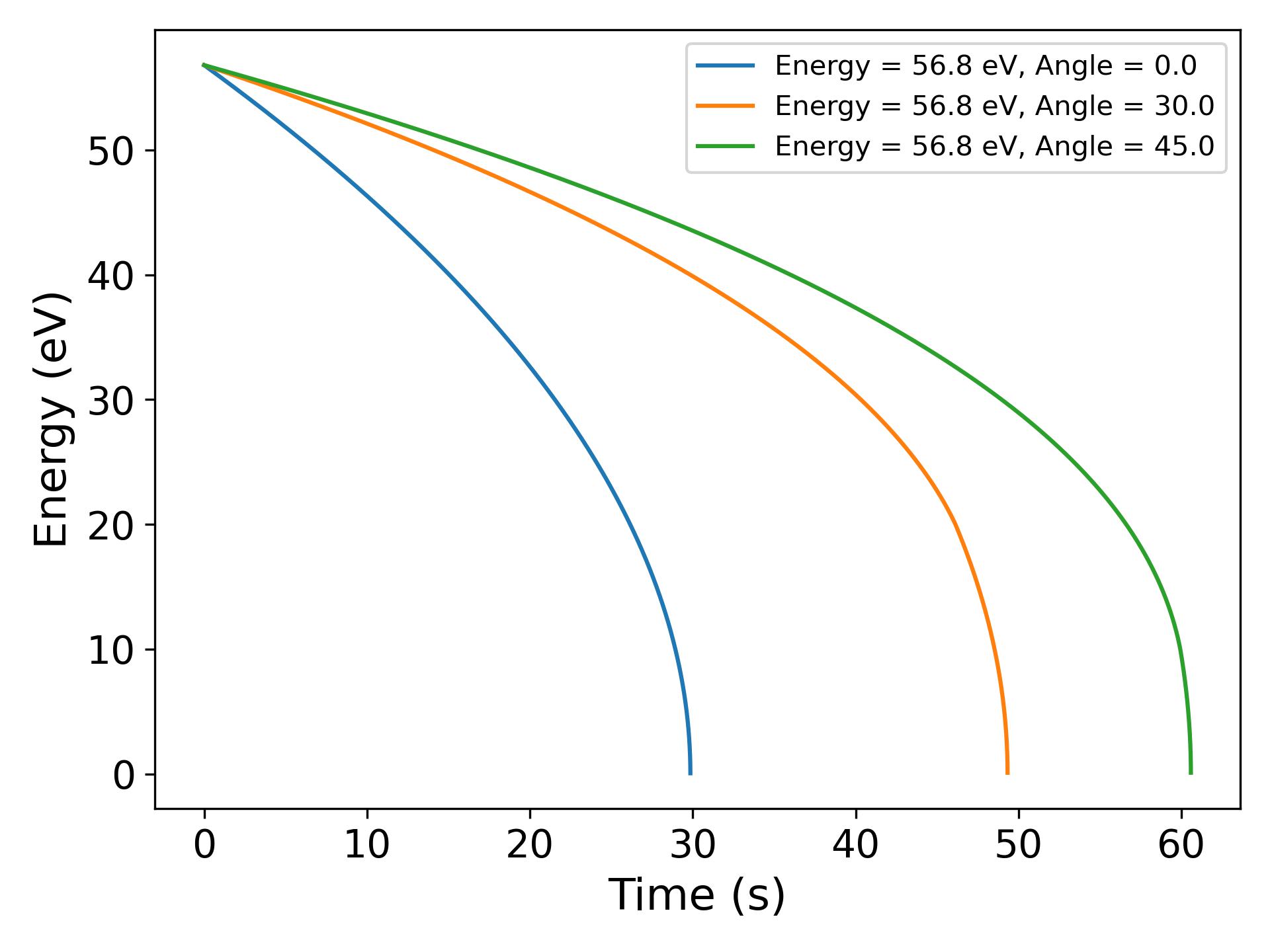}
\caption{Energy loss of a hot ion interacting with a cold cloud of ions through Coulomb collisions. The rate is non-exponential due to the energy-dependent Spitzer cooling and interaction times.} 
\label{fig:ion cooling}
\end{figure}

When the amplitude of the ion's oscillation falls below the radius of the cloud $F_z$ and $F_+$ become unity. 
Both the cooling ($\tau_{Coul}$) and the interaction times are energy-dependent. Figure \ref{fig:ion cooling} shows the sympathetic cooling of ions with several different initial angles in the trap. Due to the strong energy dependence of $\tau_{Coul}$, $F_z$, and $F_+$ the {sympathetic} cooling time varies from 10s of ms to 10s of seconds.

\subsubsection{Detection Time}

The energy exchange rate between the ion and the tank circuit is given by Equation~[\ref{cooling rate}] when the ion is perfectly on resonance. However, the cooling rate also depends on the frequency shift of the ion motion with respect to the resonant frequency of the resonator.  

\begin{equation}
    \tau_{res}(\omega) = \frac{1}{\gamma_{res}(\omega)} = \frac{mD_{eff}^2}{q^2 Re(Z(\omega))},
    \label{eq: cooling time}
\end{equation}
where $Re(Z(\omega))$, the resonator impedance, is given by

\begin{equation}
    Re(Z(\omega)) = \frac{1/R_p}{(\frac{1}{R_p})^2+\left (\omega C - \frac{1}{\omega L} \right )^2}
    \label{real_imped}
\end{equation}
with inductance $L$, capacitance $C$, resonant frequency $\omega_0 = \frac{1}{\sqrt{LC}}$ and $Q = \frac{R_p}{\omega_0 L}$. When the ion's axial frequency is tuned to the resonant frequency of the resonator ($\omega_z = \omega_0$), the effective equivalent impedance is simply $R_p$. {In the presence of space charge effects in the trap, the eigenfrequencies will shift relative to the single-particle frequency given in Section \ref{Penning trap details}: $\omega_z \rightarrow \omega_z'$. These effects have been characterized previously \cite{SpaceCharge_1989}. However, the frequency of the trap can always be tuned to bring the ions onto resonance with the tank circuit, $\omega_z' = \omega_0$.} 
In an ideal, perfectly harmonic trap with uniform magnetic field, the eigen-frequencies in the trap are ion energy-independent. However, even a small anharmonicity in the electrostatic potential well or inhomogeneity in the magnetic field will give rise to energy-dependent frequency shifts \cite{Geonium}. Shifts in the axial and {modified} cyclotron frequencies due to electrostatic and magnetic deviations from the ideal can be parameterized with expansion coefficients $C_4$, $C_6$ and $B_2$. We use reasonable estimates for the trap inhomogeneities of $C_4 = 10^{-4}$, $C_6=10^{-5}$ and $B_2 = 6 \times 10^{-6}$ $\rm T/m^2$ \cite{BASE}.  If we consider the ion to start on resonance with the resonator ($\omega_z' = \omega_0$), these deviations have the effect of shifting the frequency: $\omega_z' = \omega_0 + \Delta \omega_z$. A further discussion of the implications of frequency detunings due to space charge and anharmonicities in the trap is provided in the following section.

\begin{figure*}[t]
    \centering
    \includegraphics[width=0.325\textwidth]{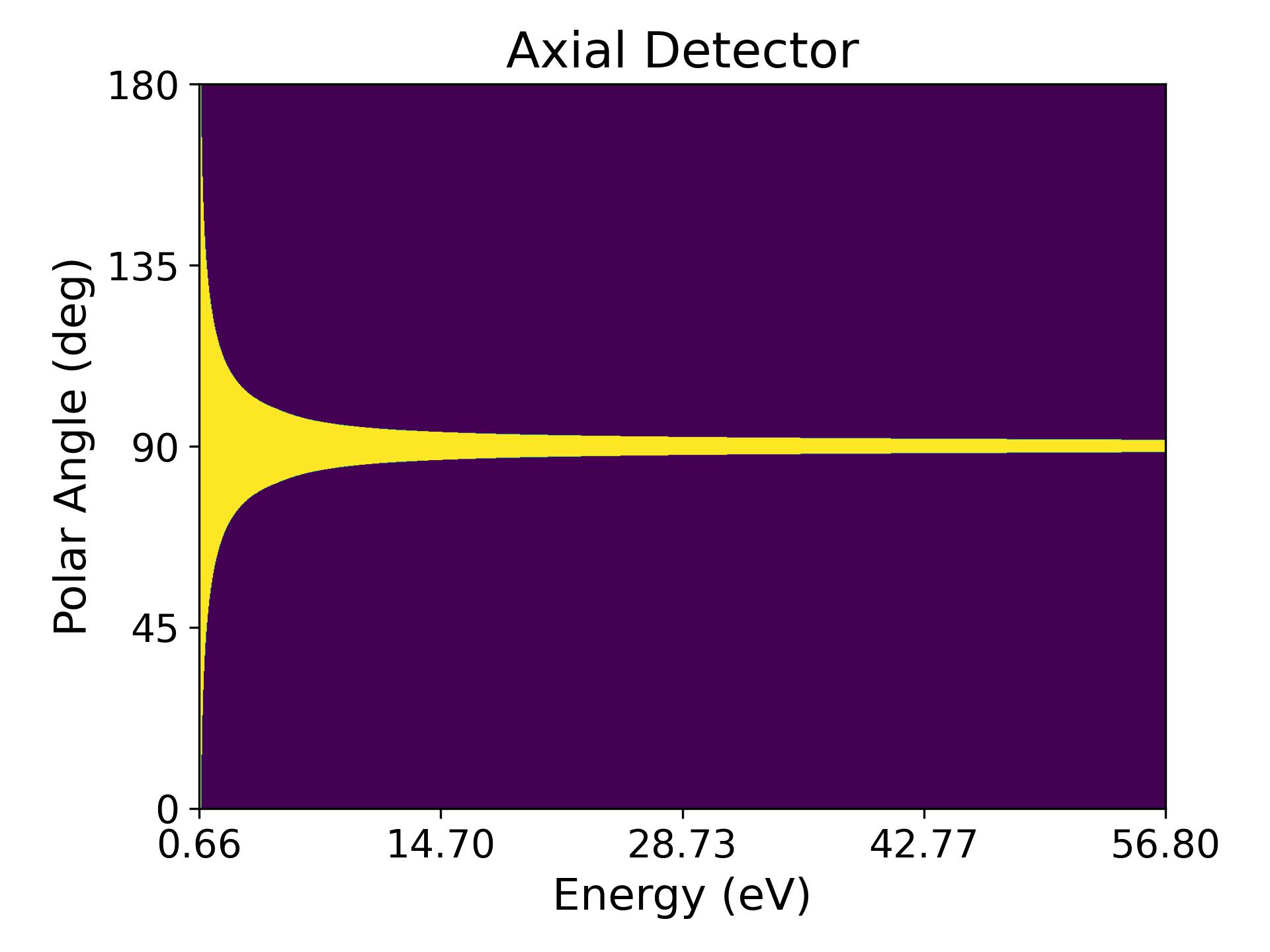}
    \includegraphics[width=0.325\textwidth]{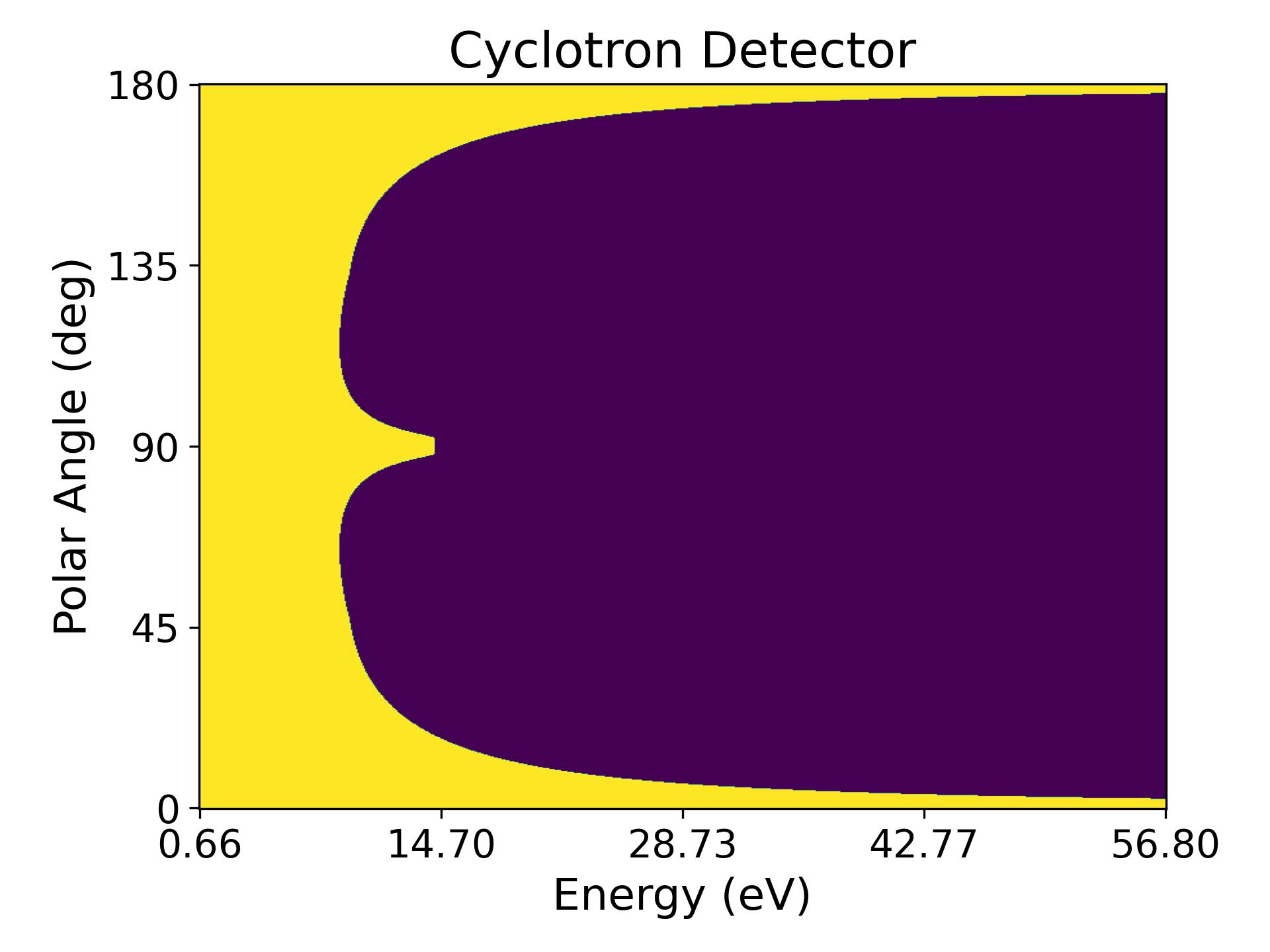}
    \includegraphics[width=0.325\textwidth]{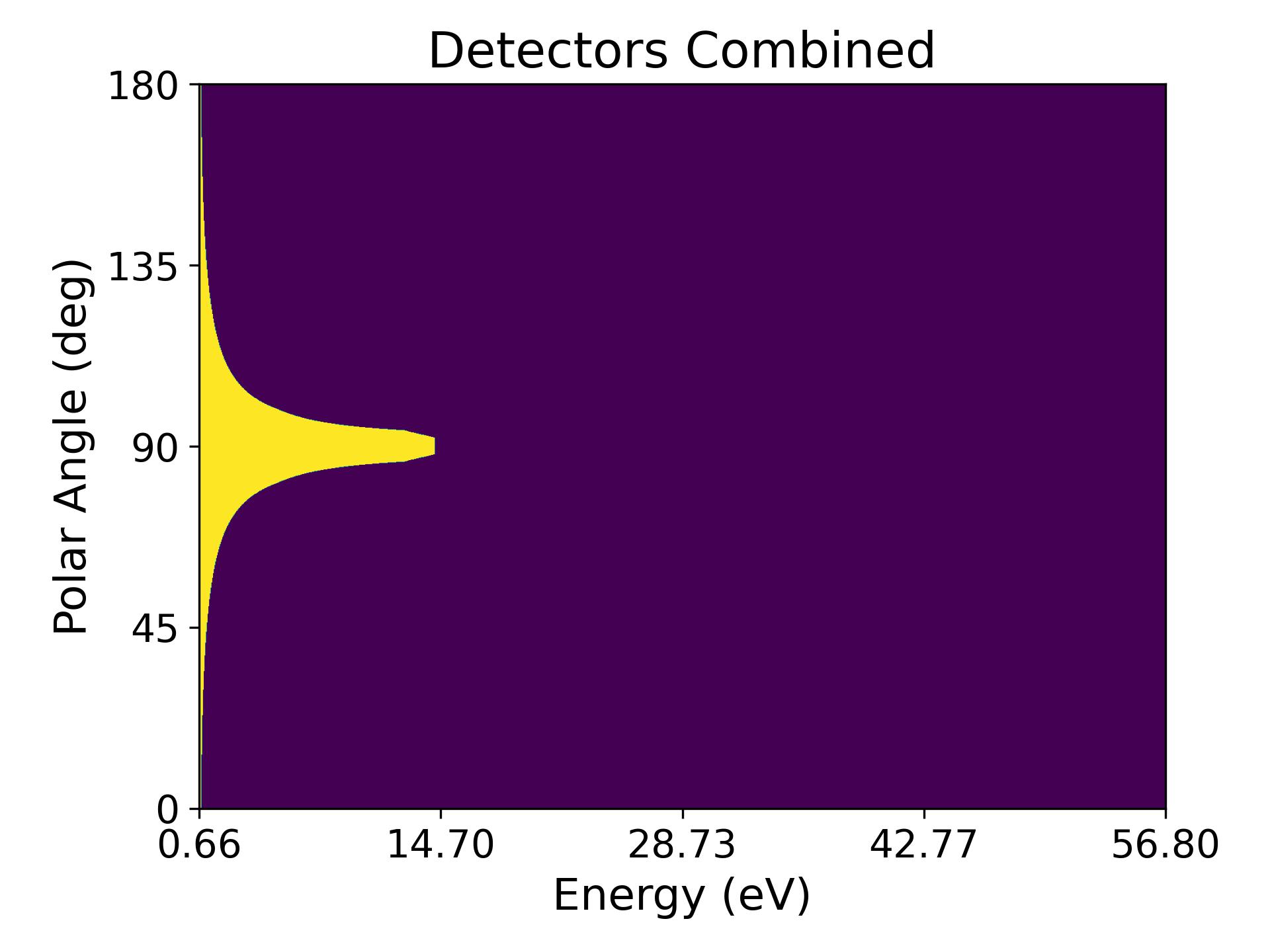}
    \caption{Left: Sensitivity of the axial resonator to decays as a function of initial angle and energy of the recoil ion in the trap. The purple and yellow regions represent initial recoil conditions the detector is and is not sensitive to, respectively. Middle: Sensitivity of the  cyclotron detector. Right: Overlap of sensitivity of both detectors. A decay is successfully detected if its initial conditions fall within the purple region.}  
    \label{fig:decay sensitivity}
\end{figure*}

{As the ion oscillates in the trap it deposits energy in the resonator. After a measuring time $\tau_{ave}$, the ratio of the signal to the Johnson noise in the axial resonator (SNR) is given by:}

\begin{equation} 
    SNR_z = \frac{1}{\sqrt{4 k_B T \tau_{ave}}} \int_0^{\tau_{ave}} \sqrt{\frac{E_z(t)}{\tau_{res}^{(z)}(\omega_z')}} dt
    \label{SNR_calc}
\end{equation}
{where $T$ is the temperature of the resonator, $E_z(t)$ is the axial ion energy as a function of time, $\tau_{res}^{(z)}(\omega_z')$ is the axial cooling time (given in Equation~[\ref{eq: cooling time}]) \cite{Dehmelt_SNR, Ulmer_thesis}. The function $E_z(t)$ accounts for the cooling of the ion due to both sympathetic cooling and resistive cooling through the resonator. The detuning with respect to the center frequency of the resonator is calculated at each time step of the integral in Equation~[\ref{SNR_calc}] to determine $\tau_{res}^{(z)}(\omega_z')$.
This holds for both the axial and {modified} cyclotron modes with the replacements $E_z \rightarrow E_+$ and $\tau_{res}^{(z)}(\omega_z') \rightarrow \tau_{res}^{(+)}(\omega_+')$. To resolve the decay of an ion on a given detector we impose a condition that there exists a $\tau_{ave}$ such that $SNR>5$. Depending on the initial conditions, averaging times to achieve this SNR are typically hundreds of milliseconds.}

\subsubsection{Decay Sensitivity}
\label{decaysensitivity}

Ninety percent of $\rm ^7Be$ decays lead to the ground state of $\rm ^7Li$, giving rise to a mono-energetic recoil energy of 56.83 eV. At this energy, the effect of the interaction of the recoil ion with the resonator dominates over the effect of sympathetic cooling.  
This is a consequence of the $E^{3/2}$ scaling of the Coulomb collision time constant in Equation~[\ref{Coulomb}]. As a result, $SNR >> 1$ on both detectors, regardless of the initial recoil angle in the trap. The detection technique is then sensitive to 100\% of these $\rm ^7Be$ decays.

In the remaining 10\% of decays through the excited state ($\rm ^7Li^*$), the daughter nucleus can have a range of energies from 0.66 to 56.83 eV. This energy is then distributed among both the axial and {modified} cyclotron modes depending on the final angle, $\theta$, with respect to the magnetic axis of the trap. {For a decay to be detected, we require $SNR_z>5$ or $SNR_+>5$. The SNR was calculated for both the axial and cyclotron detectors, {with $Q=40000$ and $Q=1000$, respectively, for all possible recoil initial conditions. The results are plotted in Figure} \ref{fig:decay sensitivity}. The yellow regions indicate the initial conditions of the recoil ions (energies and angles) that are not resolvable ($SNR<5$). As expected, the axial detector loses sensitivity when the ion recoil is perpendicular to the magnetic axis, whereas the cyclotron detector loses sensitivity when the ion decays parallel to the magnetic axis. A decay that is unresolvable on both the axial and cyclotron detectors is not resolved. The combined sensitivity is given on the right of Figure \ref{fig:decay sensitivity}.}

\subsubsection{Total Decay Sensitivity}
\label{totalsensitivity}
To calculate {the} final sensitivity to nuclear recoils, we consider the probability of the recoil ion having an initial condition inside of the purple regions of Figure \ref{fig:decay sensitivity}. The PDF for the energy of the recoil ions for decays to $\rm ^7Li^*$ is approximately uniform, as discussed in Section \ref{Decay Recoil Energy Sec}. Additionally, the final ground-state ion is emitted with an equal probability in any solid angle. For any isotropic process the PDF of emission into a polar angle is given in Equation~[\ref{eq: polar angle PDF}].

\begin{figure}[h]
\centering
\includegraphics[width=0.42\textwidth]{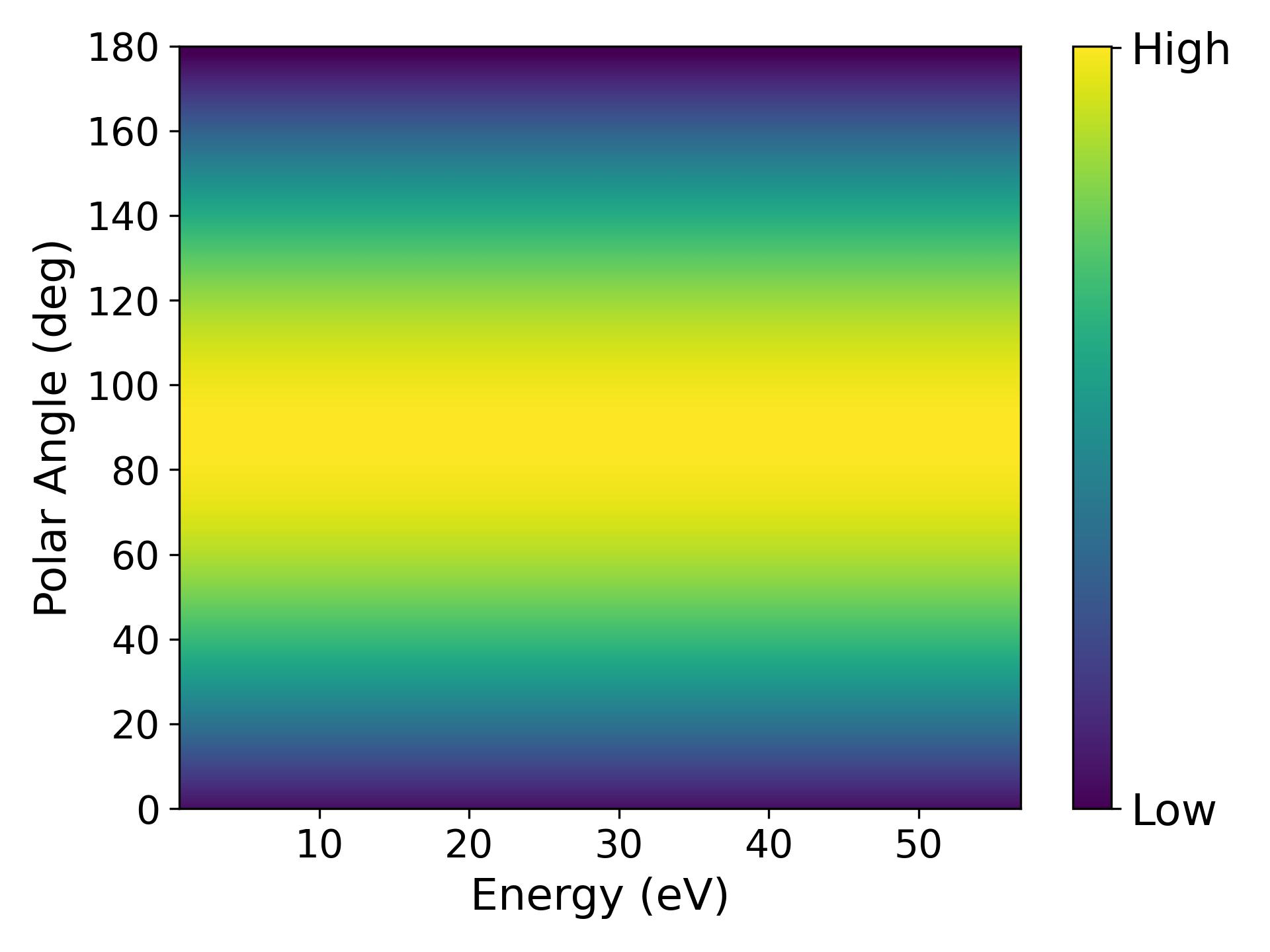}
\caption{Probability density function for the final state of $\rm ^7Li$ after decay through the excited state.} 
\label{fig:pdf}
\end{figure}

Using these two conditions we can derive the probability density function in terms of the final energy and polar angle in the trap (with respect to the magnetic axis), which is plotted in Figure \ref{fig:pdf}. Integrating the right of Figure \ref{fig:decay sensitivity}, weighted by the PDF in Figure \ref{fig:pdf}, gives a total sensitivity to decays to $\rm ^7Li^*$ of 95\%. 

\begin{figure}[H]
\centering
\includegraphics[width=0.45\textwidth]{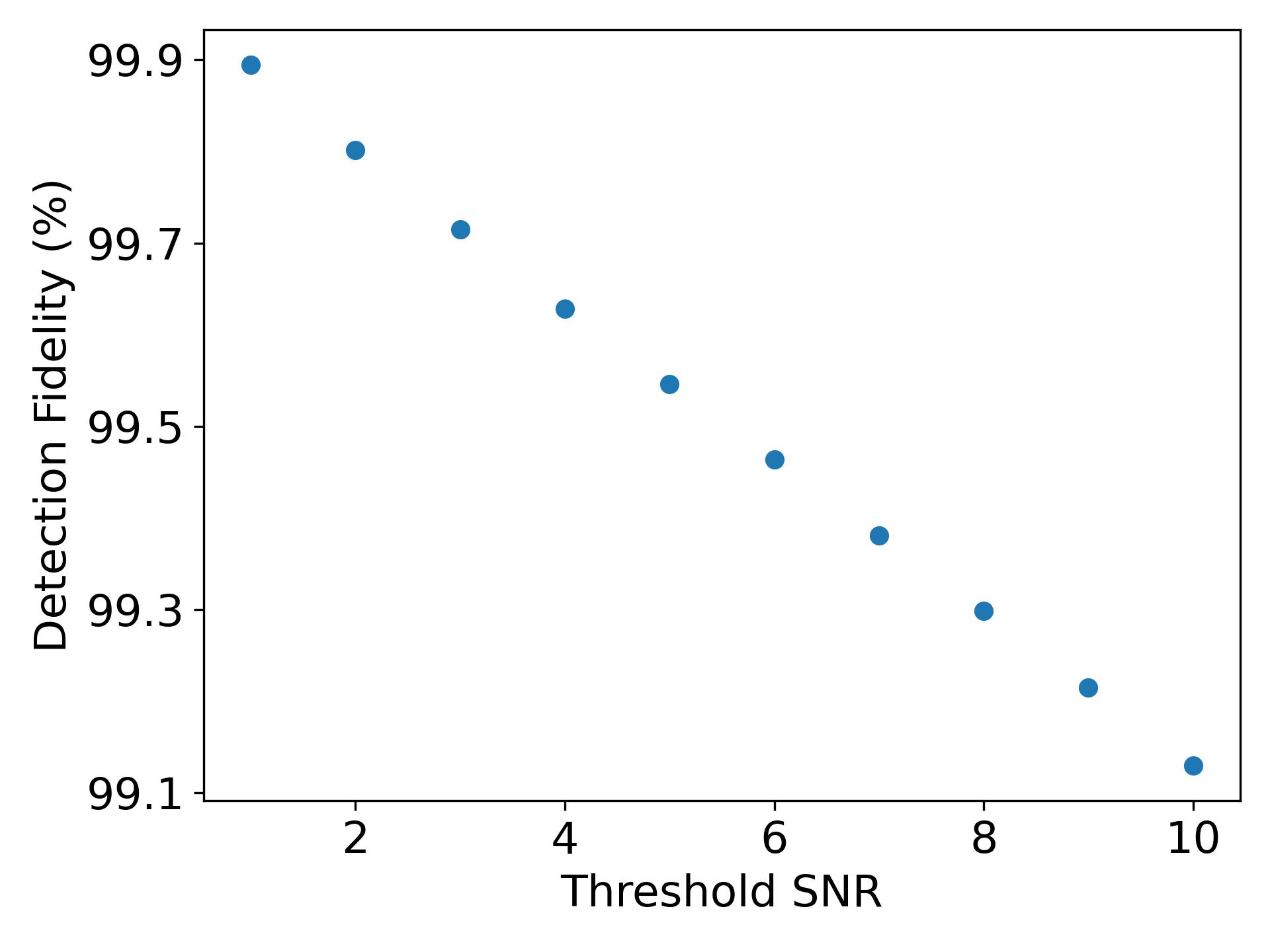}
\caption{Total detection fidelity as a function of the threshold SNR for axial resonator $Q=40000$ and cyclotron resonator $Q=1000$}
\label{fig:fidelity_vs_SNR}
\end{figure}

\begin{figure}[H]
\centering
\includegraphics[width=0.45\textwidth]{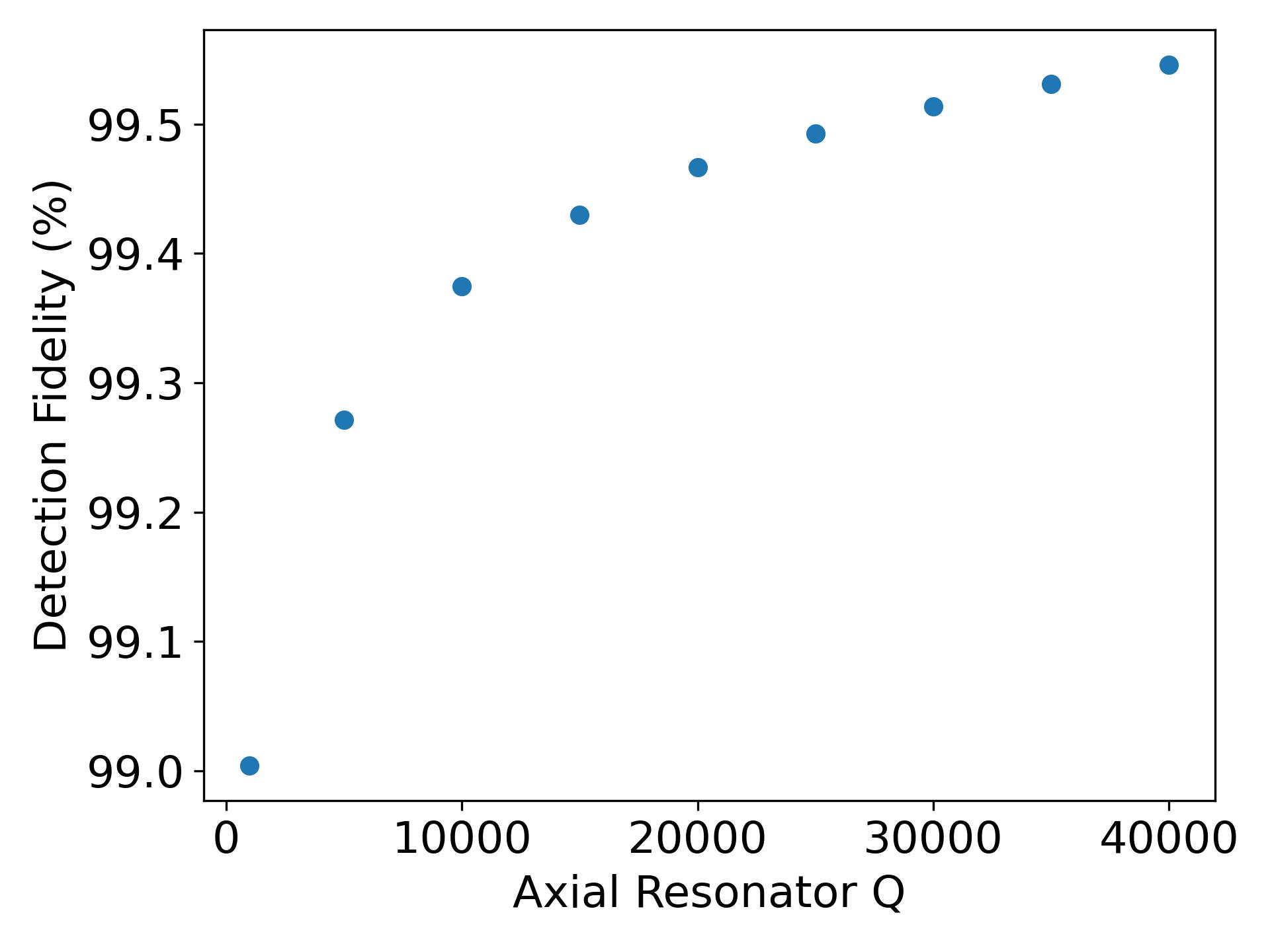}
\caption{Total detection fidelity as a function the axial resonator Q at $SNR_z>5$ and $SNR_+ > 5$ threshold}
\label{fig:fidelity_vs_Q}
\end{figure}

Adding the sensitivity to decays directly to the ground state gives a total sensitivity for the experiment of $99.5\%$. {In Figure \ref{fig:fidelity_vs_SNR}, the total sensitivity is plotted as a function of the threshold SNR, demonstrating that $>99.0 \%$ fidelity can be achieved even for a threshold of $SNR>10$. The sensitivity is plotted as a function of the $Q$-value of the axial resonator in Figure~\ref{fig:fidelity_vs_Q}, while fixing the detection threshold at $SNR>5$ for both detectors. As shown, even with $Q=1000$, a high detection fidelity of $99 \%$ is achievable. This low-$Q$ regime may be favorable as the detection is less sensitive to frequency shifts relative to the center frequency of the resonator, which may arise from space charge effects, trap anharmonicity or trap tuning errors.} Based on this analysis, it is anticipated that the proposed technique will have extremely high fidelity for the measurement of decays on an event-by-event basis.


\section{Statistical Analysis and Systematics\label{Statistical Analysis}}

In this section, we describe measurement accuracy.  
With only a single 1s electron, the expected half-life of $\rm ^7Be^{3+}$ is $\tau_{1/2} = 106$ days, and we aim for a measurement uncertainty below $5\%$. 

\subsection{Statistical Uncertainty}
\label{statun}

We assume that the original number ($N_0$) of $^7$Be$^{3+}$ ions in the trap and the amount of time between decays are measured with high enough precision that measurement error is negligible compared to the statistical error from variability inherent in the exponential decay process. The results of a  simulation presented in Section~\ref{simstudy} justify this claim.  

Let $\tau_{1/2}$ days be the true half-life of $^7$Be$^{3+}$.  
We treat time until an individual ion decays as an independent exponential random variable with parameter $\lambda = \frac{\ln(2)}{\tau _{1/2}}$.  
Let $X_i$ denote the time between the decays of ion $i-1$ and ion $i$, with $X_0=0$. 
Then $X_i$ is an exponential random variable with rate parameter $N_i\lambda$, that is, $X_i\sim \texttt{Expo}(N_i\lambda)$, where $ N_i=N_0-i+1$, and $X_i$ is independent from $X_j$ as long as $i\neq j$.  
Given observed interdecay times $\vec{x}=(x_1,x_2,\dots ,x_{N_0})$, the likelihood for $\lambda$ is given by 

\begin{equation}
    L_{\vec{x}}(\lambda)=\Pi_{i=1}^{
    N_0} N_i\lambda e^{-\lambda N_ix_i},
\end{equation}
which is maximized at $\hat{\lambda}=N_0\left(\sum\limits_{i=1}\limits^{N_0} N_ix_i\right)^{-1}$.  
The maximum likelihood estimate for $\tau_{1/2}$ is therefore $\hat{\tau} _{1/2}=\frac{\ln(2)}{\hat{\lambda}}$, with corresponding estimator 
\begin{equation}
    \frac{\ln(2)}{N_0}\sum _{i=1}^{N_0} N_iX_i ,
\end{equation}
\noindent which has mean $\frac{\ln(2)}{\lambda}=\tau _{1/2}$ and standard deviation $\frac{\tau _{1/2}}{\sqrt{N_0}}$.  
The estimator is approximately normal for large $N_0$ (greater than 20, in this context), but the exact distribution of the estimator for any $N_0$ can be simulated readily.  

Among all possible unbiased estimators, the proposed estimator is optimal in that it has the smallest possible standard deviation.  
{In other words, no accurate estimator of $\tau _{1/2}$ can be more precise, that is, there is no way to combine the data points into a better estimator without biasing the estimator or losing precision.}

We provide a proof of this statement in the Appendix~\ref{appC}.  
In Section~\ref{totalion}, we provide a detailed analysis to obtain a measurement with fractional uncertainty below 5\%.


\subsection{Total Ion Number}
\label{totalion}

If $N_0=500$, the expected amount of time until all ions to decay is 1038.8 days. The standard deviation of our estimator in this setting is $\frac{106}{\sqrt{500}}=4.74$, for a fractional uncertainty of $4.74/106 = 4.47\%$.

It may be unrealistic to wait for all $N_0$ ions to decay. Suppose instead that we wait until $D\leq N_0$ ions decay.  
The expected amount of time for $D$ ions to decay is 

\begin{equation}
    \frac{\tau _{1/2}}{\ln(2)}\sum^D_{i=1}N_i^{-1}. 
\end{equation} 



\begin{figure}[h!]
\includegraphics[width=0.48\textwidth]{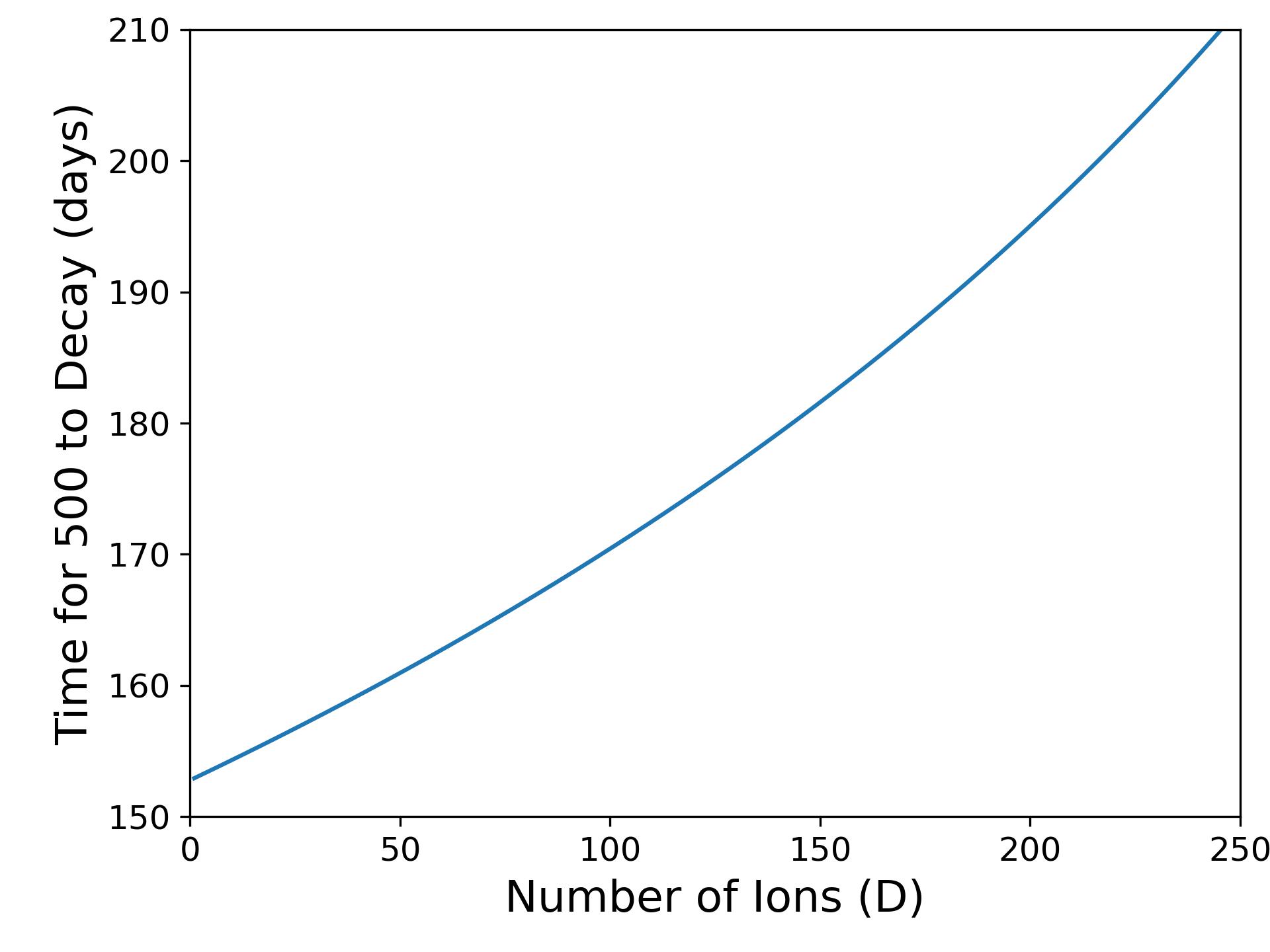}
\caption{
Expected time (in days) required for 500 ions of $^7$Be$^{3+}$ to decay beginning with 500 ions and replenishing back to 500 whenever $D$ have decayed.}
\label{fig:decayfrom500}
\end{figure}

By periodically replenishing the Penning trap, we obtain a measurement of equal precision using less time.  
The precision depends only on the number of total decays that occur: as long as 500 decays occur, the fractional uncertainty remains 4.47\%.  
If we let $D=125$ and repeat the experiment four times, replenishing to $N_0=500$ after each experimental run, then the average of the four half-life estimates has a fractional uncertainty of 4.47\%, but the expected total run time is only 175.8 days. Figure~\ref{fig:decayfrom500} illustrates the tradeoff, showing the expected run time for 500 total decays if we replenish whenever $D$ ions have decayed. 
Replenishing after every decay minimizes the expected run time to 152.93 days, but such frequent replenishing {would ultimately reduce the detection sensitivity due to the additional space charge.}
 
\subsection{Error Propagation in the Estimation of half-life}
\label{simstudy}

Sources of measurement error in this estimation process are (1) uncertainty in the initial number of ions in the trap, (2) uncertainty in registering a decay, and (3) uncertainty in the measured time between decays.  

Two types of error within Source (1) are charge exchange of $^7$Be$^{3+}$ with background molecules to form $^7$Be$^{2+}$, eliminating a $^7$Be$^{3+}$ from the stored ions, and uncertainty in the initial number of $^7$Be$^{3+}$.  
For more details, see Section~\ref{5point3}.  
Charge exchange eliminates a $^7$Be$^{3+}$ from the ensemble.  
The effect that this has on the measurement of the half-life is a maximum when the collision occurs at the beginning of the experiment.  
This is equivalent to beginning with one fewer $^7$Be$^{3+}$ ion.  
Hence we include this in Source (1).  

To investigate the impact of such errors on the final estimate of the half-life $\tau_{1/2} = 106$ days, we conducted a simulation study. We took as our context the estimation of half-life by averaging the results from four measurement campaigns, each of $D=125$ decays, with replenishing. We assumed that the errors are independent. We took the initial number of ions $N_0$ for each campaign to be truly uniformly distributed on $\{495, 496, \dots , 505\}$; but a na\"ive analysis assumes $N_0=500$. At the end of Section~\ref{totalsensitivity}, we showed that the probability of measuring a given decay is 99.5$\%$.  We took each decay to be observed independently with probability $p\in\{95\%, 97.5\% ,99\% , 99.5\%\}$; but a na\"ive analysis assumes each decay is observed. {To account for Source (3)}, we took the measured time between decays to be normally distributed about the true exponential time, with a standard deviation of 30 seconds (conservatively), truncated to be positive; but a na\"ive analysis assumes these times are exponential. 

We simulated a data set of decays four times and averaged the half-life estimates. Let $\hat{\tau}$ be the ideal (oracle) {estimator} obtained from knowing: the true value of $N_0$, the first $D=125$ decays (observed or not), and the actual interarrival times for those decays (without truncated normal measurement error). Let $\tilde{\tau}$ be {a na\"ive estimator}, the one obtained using: $N_0=500$, the first $D=125$ observed decays, and the interarrival times subject to measurement error. We repeat this entire process $100,000$ times and report the empirical means and standard deviations of our estimates for each value of $p$ in Table~1. 

\begin{table}
 \caption{Decay data simulation results. {Columns two and three describe the ideal estimator; Columns four and five describe a na\"ive one.}\label{simresults}}
 \begin{ruledtabular}
 \begin{tabular}{|c|c|c|c|c|}
 \hline
 $p$ & {avg$|\hat{\tau} |$} & {sd$|\hat{\tau} |$} & {avg$|\tilde{\tau} |$} & {sd$|\tilde{\tau} |$} \\ 
 \hline
 95\% & 106.02 & 4.74 & 112.51 & 5.09 \\  
 \hline
 97.5\% & 105.99 & 4.74 & 109.14 & 4.92 \\  
 \hline
 99\% & 105.98 & 4.75 & 107.22 & 4.83 \\  
 \hline
 99.5\% & 106.00 & 4.73 & 106.62 & 4.77 \\  
 \hline
 \end{tabular}
 \end{ruledtabular}
 \end{table}

We see that the {na\"ive} estimator is slightly positively biased, as its average exceeds 106, and it has a slightly higher standard deviation than does the oracle estimator. Both the bias and standard deviation decrease as the probability of observing a decay increases. For a probability of 99.5\%, our best estimate of the true value, the bias is 0.62 days and the standard deviation is 0.04 days more than the oracle estimator’s. In conclusion, the three sources of error are negligible. 


\section{Experimental Realization\label{Experimental Realization}} 

In previous sections, we examined the factors in an experiment to measure the half-life of $\rm ^7Be^{3+}$ using nuclear recoil to track the decay. This section is a discussion of the practical production and trapping of $\rm ^7Be^{3+}$. 

\subsection{Ion Production}

$\rm ^7Be$ can be produced by a number of nuclear reactions, such as {$\rm p+{^{10}B} \rightarrow {^7Be} + \alpha$}, with $>1$ MeV proton beams. More readily, $\rm {^7Be}$ is produced by proton spallation of oxygen in water followed by chemical separation and evaporation. $\rm ^7Be$ can be vaporized from a surface by laser ablation that produces a plasma of $\rm ^7Be$ consisting dominantly of neutrals and $\rm ^7Be^{1+}$. An Electron Beam Ion Trap (EBIT) \cite{Heidelberg_EBIT, Micke_EBIT}, can be used to produce up to fully ionized $\rm ^7Be$. An example, a compact, room-temperature EBIT that uses a high current electron beam to produce ions by electron-impact ionization, is in~\cite{Ariana's_EBIT_paper}. A unitary Penning trap, a trap with permanent magnet electrodes to produce both the magnetic and electric fields, captures the resulting ions before injection into a low-energy (2-3 keV/q) beamline. The use of a room-temperature EBIT enables the $\rm ^7Be$ source target to be replaced without breaking vacuum.

The ions are then focused through a Wien filter to select the desired ion state, in this case $\rm ^7Be^{3+}$. Though the resolving power of the Wien filter is insufficient to filter contaminant $\rm ^7Li^{3+}$ ions produced from the EBIT, these can be filtered later in the trap or in the beamline by using a Multi-Reflection Time of Flight (MR-ToF) Mass Spectrometer or in the $^7$Be$^{3+}$ Measurement Trap.

\begin{figure*}[t]
    \centering
    \includegraphics[width=0.9\textwidth]{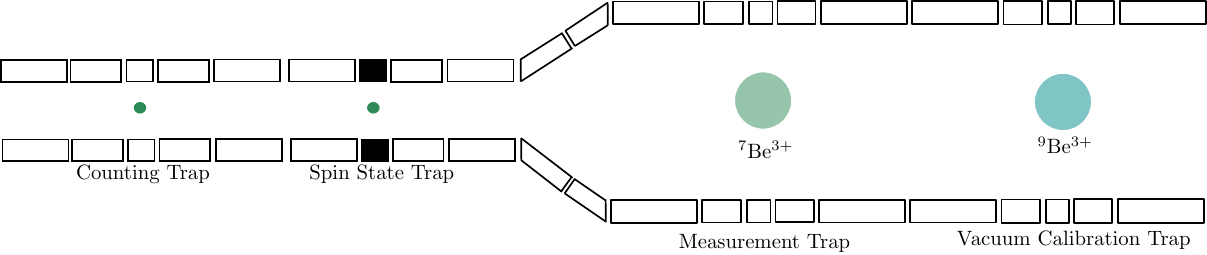}
    \caption{Penning trap structure showing a Counting Trap designed for single-ion sensitivity with tens of ions, a Measurement Trap where the $\rm ^7Be^{3+}$ is stored for the half-life measurement, and a Vacuum Calibration Trap for storing $\rm ^9Be^{3+}$.}
    \label{fig:trap structure}
\end{figure*}

\subsection{Ensemble Preparation: Trapping and Cooling} \label{Ensemble Preparation}

After ionization and filtering, the $\rm ^7Be^{3+}$ ions are injected into {a 3 T superconducting magnet that houses the Penning trap electrodes (see Figure \ref{fig:trap structure})}. A small ensemble (10s of ions) is first decelerated and trapped in a Counting Trap, where any remaining contaminant ions are resonantly ejected. The remaining $\rm ^7Be^{3+}$ can be counted {with single particle precision} through the dip width in the resonator, as discussed in Section \ref{Half-Life Measurements}. These ions are then transported to the Measurement Trap depicted in Figure~\ref{fig:trap structure}, with a large diameter and deep electrostatic potential well to ensure that recoiled $\rm ^7Li^{3+}$ ions remain trapped for detection. This loading process can be repeated until the Measurement Trap has the desired number of initial ions, $ N_0$. This particle-stacking procedure produces a large, pure ensemble in the Measurement Trap with low uncertainty on the total initial number of $\rm ^7Be^{3+}$.

\subsection{Recombination and Ion Storage Time}\label{5point3}

To measure a nuclear half-life of $\tau_{1/2}=106$ days with a small ensemble relies on all other ion-loss mechanisms, such as charge exchange ($\rm ^7Be^{3+} + e^- \rightarrow  {^7Be^{2+}}$), to have characteristic half-lives much longer than 106 days. Using a simple model for ion-neutral collisions in the trap the lifetime of a charge state is estimated by:

\begin{equation}
    \tau_{CE} = \frac{1}{\sigma p} \sqrt{\frac{k_B T \mu_m}{3}}
\end{equation}
where $\mu_m = \frac{m_Rm}{m_R+m}$, with $m_R$ being the mass of the residual gas molecule, $m$ the mass of the ion, $T$ the temperature, and $\sigma$ being the cross section of the interaction.

To satisfy the condition that $\tau_{CE} >> \tau_{1/2}$ we require that the vacuum be maintained at well below $10^{-15}$ mbar. A critical advantage of the Penning trap, compared to a storage ring, is the ability to achieve extreme high vacuum conditions. The Penning trap is contained in a cryogenic (4 K) vacuum chamber that can be isolated from the room-temperature beam line with a cryogenic valve. This eliminates backstreaming and, as shown in \cite{ALPHATRAP}, enables a vacuum better than $10^{-18}$~mbar to be achieved. The vacuum conditions in the trap can be monitored {\it in situ} by loading an ensemble of non-radioactive $\rm ^9Be^{3+}$ in a neighboring trap (Vacuum Calibration Trap in Figure \ref{fig:trap structure}). {These ions are loaded using the same production and loading mechanism for $\rm ^7Be^{3+}$}. 

Based on this pressure, in an ensemble of 500 ions, the expected number of ions that will be lost due to charge exchange with a background gas during the measurement periods presented in Section \ref{Statistical Analysis} is less than one. These losses would amount to changing the number of undecayed $\rm ^7Be^{3+}$ ions at a random point during the experiment. This systematic uncertainty is considered in Section \ref{Statistical Analysis} and demonstrated to be small. It can be reduced further by emptying the Measurement Trap back into the Counting Trap at the end of a measurement cycle and counting the number of $\rm ^7Li^{3+}$ and $\rm ^7Be^{3+}$. If the number of $\rm ^7Li^{3+}$ is consistent with the measured number of decays and the total number of ions is consistent with the initial number of ions in $t_0$, then it is known that no ions were lost due to charge exchange.

\section{{Hyperfine Branching Ratios}
\label{Hyperfine section}}
{Conservation of angular momentum requires that the decay of $\rm^7Be^{3+}$ depend on the ion's initial hyperfine state, as discussed in Section~\ref{sec: atomic structure}. As initially theorized in \cite{Folan}, if an ensemble of ions are prepared in the ground $F=2$ state, they can only decay to the ground state of $\rm ^7Li$, and no gamma rays will be emitted from the sample. A capability that is unique to the Penning trap, compared to a storage ring, is the ability to prepare an ensemble of $\rm ^7Be^{3+}$ in a pure hyperfine level: $\left |F, m_F \right >$. This is achievable by use of the continuous Stern-Gerlach effect, where an inhomogeneous magnetic field is applied to the trap causing a change in the $z$ projection of the bound electron spin to induce a measurable change in the ion's axial frequency in the trap \cite{Geonium}:}

{The spin flip transition can be driven directly with microwaves between the Zeeman sub-levels of the hyperfine states discussed in Section \ref{Motivation} and as recently demonstrated with $\rm ^9Be^{3+}$ in \cite{Dickopf}. We propose applying this technique, ion by ion, to prepare an ensemble in a pure $F=2$ hyperfine state. This can be achieved with an additional trap, denoted as the Spin-State Trap in Figure \ref{fig:trap structure}.}

{In decays directly to the ground state of $\rm ^7Li$, the recoil energy is 56.83 eV. The number of ions at this energy can be directly measured by kinematic reconstruction of the recoil. For a fixed averaging time, $\tau_{ave}$, the SNR on the axial ($SNR_z$) and cyclotron ($SNR_+$) resonators is proportional to the initial ion energy, as shown in Equation~[\ref{SNR_calc}]. For additional details, see Appendix \ref{appD}. The uncertainty in the energy measurement is limited by ions that decay with a small angle, $\theta = \delta$, with respect to the magnetic axis of the trap. In this case, only a small amount of energy couples to the cyclotron mode and is unresolvable on that detector. This decay is indistinguishable from a lower energy recoil that arose from a decay through the excited state $\rm ^7Li^*$, with $\theta = 0$. Similarly, from the exclusion plots in Figure \ref{fig:decay sensitivity}, it is  clear that lower energy decays ($<<56.83$ eV) may not be resolved on both detectors for many angles in the trap, limiting the ability to reconstruct their decay energy. For example, if an ion is measured to have an energy $E=5$ eV in the axial mode, then this may arise from a 5 eV recoil at $\theta = 0^{\circ}$ or a 10 eV recoil at $\theta = 45 ^{\circ}$. Since both decays in this case are unresolved on the cyclotron detector, this gives rise to a large measurement uncertainty of $\approx 100 \%$. Nonetheless, even for this low energy recoil, it is still possible to determine definitively that the decay did not proceed directly to the ground state with a recoil of 56.83 eV.} 

{Therefore, based on Figure \ref{fig:decay sensitivity}, we conclude that low energy recoils will have large uncertainty (in excess of $100 \%$), and high energy recoils ($>15$ eV) can be resolved with an error $<20\%$, conservatively. In fact, assuming an energy measurement uncertainty of $10 \%$, it is possible to reliably test whether the decays occur through the excited state or directly to the ground state with as few as $n=100$ measured decays. We simulate data under two hypotheses, each of which assumes independent normal measurement errors, $\epsilon \sim N(0,\sigma^2)$ (units is eV). Under the ``mixed" hypothesis ($F=1$ and $F=2$ populated according to their statistical weights), there is a 10.4\% probability that the decay energy is $\text{Uniform}(0.66, 56.83) + \epsilon$ and an  89.6\% probability that the decay energy is simply $56.83+\epsilon$. Under the ``pure" (100\% $F=2$ population) hypothesis, the decay energy is $56.83+\epsilon$. In order to give a conservative estimate, we assume, based on Figure~\ref{fig:decay sensitivity}, that any measurements below 15 eV go unobserved.} 

{The optimal test for deciding between these hypotheses is the Likelihood Ratio Test, which calculates the probability of an observed data set under each hypothesis and chooses the hypothesis associated with the greater probability \cite{casella2001}. For each combination of $n\in\{25, 50, 100, 200\}$, the number of measured decays, and $\sigma\in\{2.84, 5.68, 11.36\} $ eV (i.e., 5\%, 10\%, and 20\% of 56.83 eV), we simulated 20,000 data sets---half under the mixed model and half under the pure model. The results are presented in Table \ref{tab:sigma_table}.}

{The false positive rate (also known as the Type I error rate), is the probability that the test chooses the pure hypothesis when the data actually come from the mixed model. Ideally, this number is close to 0\%. The true positive rate (also known as power), is the probability that the test chooses the pure hypothesis when the data actually come from the pure model. Ideally, this number is close to 100\%.}

\begin{table}[h]
  \caption{{False positive and true positive rates for different combinations of $\sigma$ and $n$. Rates are rounded to the nearest tenth of a percent.}}

  \centering
  {
  \begin{tabular}{|c|c|c|c|c|}
    \hline
        & $n=25$ & $n=50$ & $n=100$ & $n=200$ \\ \hline
    &\multicolumn{4}{|c|}{$\sigma = 2.8415$ (5\%)} \\ \hline
    False positive rate & 19.5\% & 4.5\% & 0.4\% & 0.0\% \\ \hline
    True positive rate & 99.1\% & 99.7\% & 100.0\% & 100.0\% \\ \hline
   &\multicolumn{4}{|c|}{$\sigma = 5.683$ (10\%)} \\ \hline
    False positive rate & 29.2\% & 10.2\% & 1.8\% & 0.1\% \\ \hline
    True positive rate & 96.6\% & 98.3\% & 99.7\% & 100.0\% \\ \hline
   &\multicolumn{4}{|c|}{$\sigma = 11.366$ (20\%)} \\ \hline
    False positive rate & 43.0\% & 27.8\% & 16.8\% & 7.1\% \\ \hline
    True positive rate & 83.1\% & 85.8\% & 90.2\% & 95.2\% \\ \hline
  \end{tabular}
 }
  \label{tab:sigma_table}
\end{table}

{If the relative measurement error is 5\% ($\sigma = 2.84$ eV), the test is reliable with $n$ as small as $50$. Under 10\% measurement error on the energy, $n=100$ is sufficient. In the extreme case of 20\% error, with $n=200$ we can still show whether ions decay directly to ground state.}

\section{Discussion\label{Discussion}}  

In this section we discuss possible extensions of this work both with $\rm ^7Be$ directly and other electron-capture isotopes.

The measurement of the half-life of $^7$Be$^+$ probes the role of $2s$ electrons in the decay of the nucleus. 
$\rm ^7Be^{1+}$ is laser coolable by driving the $2s-2p$ transition, reducing the temperature of the ions far below the noise floor of the resonator. Further, as a singly-charged ion, the space-charge effects discussed in Section \ref{Space Charge Section} are reduced, allowing for larger numbers of ions to be trapped. A recoil measurement in this species could also be paired with the use of florescence, as demonstrated with single ion sensitivity in $\rm ^9Be^+$ \cite{Niemann_2020}.

For heavier highly-charged isotopes, the detection of the recoil in the trap is more challenging as the cooling time constant due to Coulomb collisions scales as $q^{-4}$ and the averaging time scales as $q^{-2}$. This can be compensated by reducing the number of stored ions during a measurement cycle, but will then require longer data collection time. This can be compensated for by using more than one trap in parallel during a given measurement cycle. Nonetheless, these isotopes are particularly interesting due to the short hyperfine state lifetimes.

\begin{table}[h] 
 \caption{Heavier electron-capture radioisotopes for measurement in Penning ion trap\label{Heavier Isotopes}}
 \begin{ruledtabular}
 \begin{tabular}{||c c c c c||}
 \hline
  & & \textbf{H-like} & \textbf{Recoil} & \textbf{Hyperfine}\\
 \textbf{Nucleus} & $\tau _{1/2}$ & \textbf{Charge} & \textbf{Energy} & \textbf{State} \\ 
  & & \textbf{state} & \textbf{(eV)} & \textbf{Lifetime} \\ [0.5ex] 
 \hline\hline
 $\rm ^{37}Ar $ & 35\text{ days} & 17 & 9.62 & 10 h \\ 
\hline
 $\rm ^{49}V$ & 337\text{ days} & 22 & 3.97 & 193 s\\
 \hline
 $\rm ^{51}Cr $ & 27.7\text{ days} & 23 & 5.96 & - \\
 \hline
$\rm ^{131}Cs $ & 9.7\text{ days} & 54 & 0.51 & 31 ms \\
 \hline
 \end{tabular}
 \end{ruledtabular}
\end{table}

This also presents the possibility of preparing an ensemble in a pure hyperfine state. A few such isotopes are given in Table 2.

\section{Conclusion}

{In this manuscript, we propose a new method for measuring the half-life of highly charged radioisotopes by non-destructive nuclear recoil detection in a Penning trap. This technique enables measurements of isotopes that are difficult to access by current techniques in storage rings. Specifically, we emphasize the case of $\rm ^7Be$ that has an important role in stellar evolution and the production of solar neutrinos. By simulating the sympathetic cooling of the daughter recoil nucleus ($\rm ^7Li$) with the cold cloud of $\rm ^7Be$ we demonstrate a 99.5\% detection fidelity by non-destructive image current detection. We present a statistical analysis of half-life measurements in ensembles containing only hundreds of ions, and demonstrate that a statistical uncertainty of $<5 \%$ is achievable with 500 measured decays. The proposed non-destructive measurement technique can be extended to fully reconstruct the energy of the daughter nucleus, enabling the ability to measure branching ratios from different hyperfine states. In particular, we present the feasibility of pairing this detection scheme with developed techniques to control hyperfine states in light ions to prepare an ensemble in a pure hyperfine state, thereby directly manipulating their possible radioactive decay modes.}

\section{Acknowledgements}
{The authors thank X. Fan, A. Brinson, R.F. Garcia Ruiz, and L. Orozco for the useful discussions. SBM acknowledges support from the National Science Foundation Graduate Fellowship and the Fannie and John Hertz Foundation Fellowship. A.S. acknowledges support from a National Science Foundation Graduate Fellowship.}

\section{Author Declarations}

The views expressed in this article are those of the authors and do not reflect the official policy or position of the U.S. Naval Academy, Department of the Navy, the Department of Defense, or the U.S. Government. {This material is based upon work supported by the National Science Foundation Graduate Research Fellowship Program under Grant No. DGE 2236417. Any opinions, findings, and conclusions or recommendations expressed in this material are those of the author(s) and do not necessarily reflect the views of the National Science Foundation}

\bibliography{formatted}
\bibliographystyle{unsrt}

\appendix

\section{Derivation of Recoil Energy}

If we consider the $\rm ^7Li^{*}$ to be moving along the $z$ axis with momentum $p_{e}$, then the emitted gamma ray, with momentum $ p_\gamma$ at polar angle $\theta$ from $\hat{z}$ will modify the momentum ($ p_g$) of the final, ground-state daughter ($\rm ^7Li$).
\begin{equation}
    p_g(\theta) = \sqrt{(p_{e}-p_\gamma(\theta) \cos\theta)^2 + (p_\gamma(\theta) \sin\theta)^2}
    \label{eqn:p_g}
\end{equation}
Here $p_\gamma$ is given in the lab-frame, which is Doppler shifted compared to its momentum when emitted from the isomer at rest, and $\theta$ is the polar angle of the $\gamma$, measured from the $z$-axis. 
This Doppler shift is dependent on the angle the gamma ray is emitted with respect to the momentum of the isomer as follows  
\begin{equation}
    p_{\gamma}(\theta) = p_{\gamma0} \left( 1 + \frac{p_{e}}{m_{e}c} \cos\theta \right)
    \label{eqn:p_gamma}
\end{equation}
where $p_{\gamma0}$ is the momentum of the gamma in the rest frame of the recoiled ion~\cite{BEEST_thesis, Doppler}. From these we can compute the final recoil energy of the ground-state $\rm ^7Li$, shown in Figure \ref{fig:excited state recoil}. As seen, the decay through the excited state of $\rm ^7Li$ gives rise a range of final-state recoil energies from 0.66 to 56.83 eV.

\section{Proof that the Recoil Energy Distribution is Approximately Uniform\label{appB}}

\begin{figure}[h!]
\centering
        \includegraphics[width=0.48\textwidth]{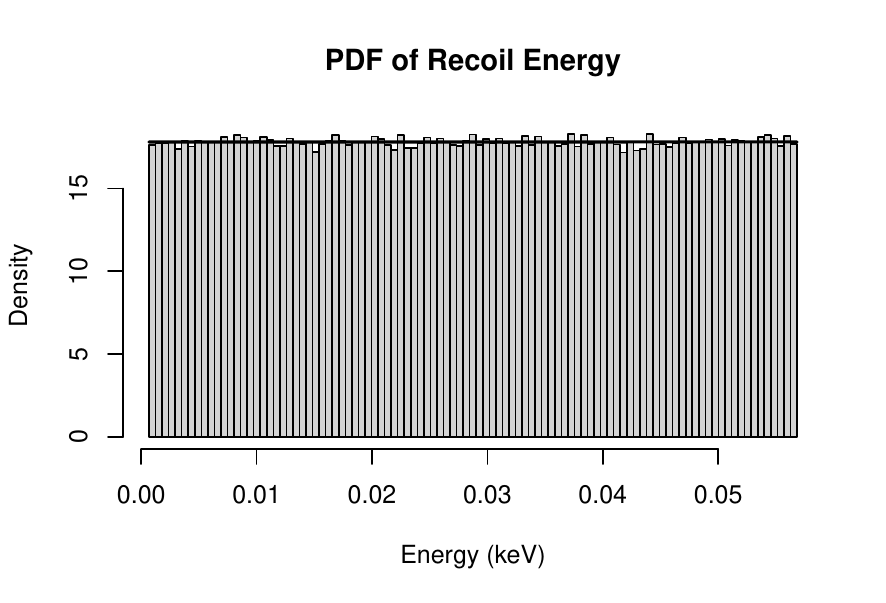}
        \caption{Histogram of $10^6$ random energies, with corresponding PDF $f_Y$ (solid black line). The induced distribution is approximately uniform.}
        \label{fig:UniformPDF}
    \end{figure}

Recall equations (\ref{eqn:p_gamma}) and (\ref{eqn:p_g}), reproduced here for convenience, along with the equation for the recoil energy as a function of polar angle $\theta$:
\begin{eqnarray*}
    \rm p_{\gamma}(\theta) &=& p_{\gamma0} \left( 1 + \frac{p_{e}}{m_{e}c} \cos\theta \right), \\
    \rm p_g(\theta) &=& \sqrt{(p_{e}-p_\gamma(\theta) \cos\theta)^2 + (p_\gamma(\theta) \sin\theta)^2}, \\
     E_g(\theta) &=& \sqrt{p_g(\theta)^2 + m_g^2c^4} - m_gc^2.
\end{eqnarray*}
\noindent We take $c=1$ and use the following values for the other constants (all measured in keV): 
\begin{eqnarray*}
    m_g &=& 7.01600343666\times931494.0954, \\
    m_e &=& 6.53584339 \times 10^6, \\
     p_{\gamma 0} &=& 477.6035, \\ 
     p_e &=& 384.2752032.
\end{eqnarray*}
Let $X$ be a random polar angle having PDF $f_X(\theta) = \frac12\sin(\theta)$, and let $Y=E_g(X)$ be the associated recoil energy. The PDF of $Y$, which we will denote $f_Y$, is given by
$f_Y(y) = f_X(E_g^{-1}(y))| \frac{d}{dy}E_g^{-1}(y)|$. Using a computer algebra program, we find that for energy $y\in (0.0006663529, 0.05683161)$ keV,
\begin{equation*}
f_Y(y) = \frac{y-k_4}{\sqrt{4k_2((y-k_4)^2-k_1)+k_3^2}},
\end{equation*}
where 
\begin{eqnarray*}
    k_1 &=& p_e^2+ p_{\gamma 0}^2+m_g^2, \\
    k_2 &=& p_{\gamma 0}^2(\frac{p_e}{m_e})^2-2p_e p_{\gamma 0} (\frac{p_e}{m_e}), \\
    k_3 &=& 2p_{\gamma 0}(p_{\gamma 0}(\frac{p_e}{m_e})-p_e), \\
    k_4 &=& -m_g.
\end{eqnarray*}
We confirmed these results with a Monte Carlo simulation of $10^6$ angles generated according to $f_X$, each of which was converted into an energy via the $E_g$ function. Figure \ref{fig:UniformPDF} is a histogram of simulated energies, which matches the theoretical $f_Y$ perfectly. We see that the distribution is approximately uniform, as claimed. 

\section{Proof that the Proposed Estimator is Optimal\label{appC}}

We let $N=N_0$ and $\tau=\tau_{1/2}$ for ease of notation.  Our maximum likelihood estimator for $\tau$ is equal to $\frac{\ln(2)}{N}\sum_{i=1}^N N_iX_i$. Observe that $X_iN_i \sim \texttt{Expo}(\lambda)$ are identically distributed independent random variables. Thus, their empirical mean $\frac{1}{N}\sum_{i=1}^N N_iX_i$ is asymptotically normal by the Central Limit Theorem. The mean and variance of the estimator for any finite $N$ are $\frac{\ln(2)}{\lambda}=\tau$ and $\tau^2/N$.    

We adjusted the notation in the following result from~\cite[Theorem~7.3.9]{casella2001} to match ours. 

\begin{theorem}[Cramer-Rao Inequality]
Let $X_{1:N}=(X_1, \ldots, X_N)$ be a sample with PDF $f(\vec{x}\mid\tau)$, and let $W(X_{1:N})$ be any estimator satisfying:
$\frac{d}{d\tau} E[W(X_{1:N})] = \int \frac{\partial}{\partial \tau}[W(\vec{x})f(\vec{x}\mid\tau)]\, d\vec{x}$
and $Var[W(X_{1:N})] < \infty$. Then 
\begin{eqnarray}
\label{CRLB}
Var[W(X_{1:N})] &\geq& \frac{(\frac{d}{d\tau}E[W(X_{1:N})]^2}{E[(\frac{\partial}{\partial\tau}\ln f(X_{1:N}\mid\tau))^2]}.
\end{eqnarray}
\end{theorem}

If we let $W$ be an estimator which is unbiased for $\tau$, then $\frac{d}{d\tau} E[W(X_{1:N})]=1$ and the numerator on the right side of Inequality (\ref{CRLB}) is 1. 
To calculate the denominator, let $k=\ln(2)$ and $\tau=k/\lambda$.  
We obtain the following.  
\begin{eqnarray*}
\ln f(\vec{X}\mid\tau) &=& \sum_{i=1}^N  \ln(N_i) + \ln(k) - \ln(\tau)- k N_i X_i \tau^{-1}  \\
 \frac{\partial}{\partial\tau} \ln f(\vec{X}\mid\tau) &=& \frac{-N}{\tau} + \frac{k\sum_i N_iX_i}{\tau^2}\\
  \left(\frac{\partial}{\partial\tau} \ln f(\vec{X}\mid\tau) \right)^2&=& \frac{N^2}{\tau^2} + \frac{k^2\left(\sum_i N_iX_i\right)^2}{\tau^4}-\frac{2Nk\sum_i N_iX_i}{\tau^3}
\end{eqnarray*}

The expected value of this is the denominator on the right side of Inequality (\ref{CRLB}). Observe that $X_iN_i \sim \texttt{Expo}(k/\tau)$ so $E[\sum_{i=1}^N X_iN_i] = N\tau/k$, $Var[\sum_{i=1}^N X_iN_i] = N \tau^2 /k^2$, and $E[(\sum_{i=1}^N X_iN_i)^2] = N\tau^2/k^2 + N^2\tau^2/k^2 = (N+N^2)(\frac{\tau^2}{k^2}).$ We now compute the expected value of the denominator in Inequality (\ref{CRLB}):
\begin{eqnarray*}
E\left[\left(\frac{\partial}{\partial\tau}\ln f(X_1, \ldots, X_N\mid\tau)\right)^2\right] &=& \\ E\left[\frac{N^2}{\tau^2} + \frac{k^2(\sum_i N_iX_i)^2}{\tau^4}-\frac{2Nk\sum_i N_iX_i}{\tau^3}\right]
&=& \\ \frac{N^2}{\tau^2} + \frac{k^2}{\tau^4}\frac{\tau^2}{k^2}(N+N^2)-\frac{2N^2k}{\tau^3}\frac{\tau}{k} 
&=& \frac{N}{\tau^2}.
\end{eqnarray*}
 
Hence, for an unbiased estimator $W$, $Var[W]\geq \tau^2/N$ and 
sd$[W] \geq \tau/\sqrt{N}$. Our estimator is unbiased since it has expected value $\tau$. Our estimator also has variance $\tau^2/N$. Because it attains this lower bound, no unbiased estimator of $\tau$ can have smaller variance than ours.

\section{Energy Calibration of the Resonator}
\label{appD}

{The exact relationship between the initial ion recoil energy and SNR on a given detector depends on the complicated dynamics of the cooling process, which governs the function $E(t)$. For a single ion interacting with the resonator this is demonstrated in \cite{Ulmer_thesis} and an energy resolution of $10 \%$  is achieved for $E \approx 100 $ meV. In the presence of sympathetic cooling, $E(t)$ can be determined through a model, such as the one presented in Section \ref{Space Charge Section} or through a calibration of the trap. Before performing the measurements in Section \ref{fig:hyperfine decay scheme}, 500 $\rm ^{12}C^{3+}$ ions can be co-trapped and cooled with a single $\rm ^7Be^{3+}$ ion. Critically, these ions will have very different oscillation frequencies in the trap but similar sympathetic cooling dynamics due to their charge states. A resonant excitation can be applied to the axial and cyclotron modes of only the single $\rm ^7Be^{3+}$ ion. This excitation can be tuned to give the ion a known initial energy. The ion will then cool back to equilibrium through Coulomb collisions with the cold $\rm ^{12}C^{3+}$ ions, resembling the conditions in the experiment described in section \ref{Space Charge Section}. The SNR on the detectors can then be directly measured for all possible different initial conditions. }

\end{document}